\documentstyle[12pt,epsf,psfig]{article}

\def\beq{\begin{equation}}
\def\eeq{\end{equation}}
\def\beqa{\begin{eqnarray}}
\def\eeqa{\end{eqnarray}}
\def\a {{\rm f}}
\def\e{r}
\def\g{\xi}
\def\w{\rho}
\def\y{\eta}

\begin{document}

\begin{flushright}
ITP-SB-98-54
\end{flushright}

\begin{center}{\bf\Large\sc Rapidity Gaps and Color Evolution in QCD Hard Scattering \footnote{Talk given at the 
International Euroconference on Quantum Chromodynamics (QCD '98),
Montpellier, France, July 2-8, 1998.} }

\vglue 1.2cm
\begin{sc}
Gianluca Oderda \\
\vglue 0.5cm
\end{sc}
{\it Institute for Theoretical Physics \\
SUNY at Stony Brook,
Stony Brook, NY 11794-3840, USA}\\
\end{center}
\vglue 1cm
\begin{abstract}
We discuss rapidity-gap events between two jets produced at high momentum transfer in $p$
$\bar{p}$ scattering, from the 
point of view of the soft energy flow into the interjet region.
We define a gap cross section and, in perturbative QCD (pQCD), resum all the leading logarithms 
in the soft intermediate energy. 
We show that the numerical result from our cross section 
reproduces the shape of the D0 and CDF \cite{D0,CDF,D0fig} experimental data.
\end{abstract}
\vfill

\newpage
\section{Introduction}
\label{sec:intr}
Dijet rapidity-gap events, identified by very low hadron multiplicity in the rapidity region between
two jets produced at high momentum transfer, have been observed 
both at Fermilab \cite{D0,CDF} and DESY \cite{ZEUS,H1}.
We refer in particular to the experimental papers of the CDF and D0 collaborations \cite{D0,CDF,D0fig}, 
where an excess of 
opposite-side dijet events with respect to a background of same-side events  
is reported in the inclusive process $p(p_A)+\bar{p}(p_B) \rightarrow J_1(p_1)+J_2(p_2)+X$, 
for low hadron multiplicity in the central region of the calorimeter detector.

This phenomenon has been originally predicted  from the 
exchange of two or more hard gluons in a color-singlet configuration,
so that color recombination between nearly opposite moving particles, and consequent interjet 
hadronization, is avoided \cite{Bjork}.

In our analysis of the problem \cite{StOd1,StOd}, we consider energy flow instead of particle multiplicity.
In these terms, we first identify a dijet rapidity-gap cross section and, 
using pQCD, write it in a factorized form that enables us 
to compute the  evolution towards long distances
of the possible color components exchanged in the partonic hard scattering.
Such evolution is due to soft, but still perturbative, interactions between the active partons.
In our analysis, gaps turn out to have a more complicated structure than just color singlet exchange \cite{eboli,zepp}.
The numerical results we obtain from  our
cross section closely resemble
the qualitative behavior of the experimental data of Ref.\ \cite{D0fig}.

\section{Dijet rapidity-gap cross section}
\label{sec:rapgapcs}
In terms of the (pseudo)rapidity variable 
$y=(1/2)\ln\cot(\theta/2)$,
with $\theta$ the polar angle, we identify  by
the condition $|y|>y_0$ the forward and backward regions of the calorimeter, where 
the two jets are to be found,  with transverse energies above the experimental threshold $E_T$ (see Fig.\ 1 of
Ref.\ \cite{StOd1}). 
We define $Q_c$ the amount of energy flowing into the symmetric central region, of width $\Delta y = 2y_0$.

The inclusive dijet cross section for all events with energy in the central region equal to $Q_c$ can be written in the 
factorized form:
\beqa
&&\frac{d\sigma}{dQ_c}\left(S,E_{T},\Delta y\right) 
=
 \sum_{f_A,f_B=u,d} \int d\cos\hat{\theta}\,
\nonumber\\ 
&& \hspace{-5mm} \times \int_0^1 dx_A \int_0^1 dx_B\;
 \phi_{f_A/p}(x_A,-\hat{t}) \,
\phi_{\bar{f}_B/\bar{p}}(x_B,-\hat{t}) \nonumber \\
&& \times \sum_{f_1,f_2=u,d}\frac{d\hat{\sigma}^{(\a)}}
{dQ_c \, d\cos\hat{\theta}} 
\left(\hat{t},\hat{s},y_{JJ},\Delta y,\alpha_s(\hat{t})\right)\, ,
\nonumber \\
\label{crosssec}
\eeqa 
where, for simplicity, we only consider the contribution of valence quarks  and  of the
partonic process $q(k_A)+\bar{q}(k_B) \rightarrow q(k_1)+\bar{q}(k_2)+X$.
We identify by $\phi_{f_A/p}$ and $\phi_{\bar{f}_B/\bar{p}}$ the  
valence parton distributions, evaluated at scale $-\hat t$,
the dijet momentum transfer. ${d\hat{\sigma}^{(\a)}}/{dQ_c \, 
d\cos\hat{\theta}}$
is a hard scattering function, starting with the Born cross section
at lowest order.  The 
index $\a$ denotes
$f_A+\bar{f}_B \rightarrow f_1 +\bar{f}_2$. 
The detector geometry determines the phase space
for the dijet total rapidity, $y_{JJ}$, 
the partonic center-of-mass (c.m.)\ energy squared, $\hat{s}$, and
the partonic c.m.\ scattering angle $\hat{\theta}$, with
$-\frac{\hat{s}}{2}\left(1-\cos \hat{\theta} \right)=\hat{t}$.

\section{The partonic cross section: factorization and evolution in color space}
\label{sec:facevoco}
The partonic cross section ${d\hat{\sigma}^{(\a)}}/{dQ_c \, 
d\cos\hat{\theta}}$ is an IR safe quantity. 
It can be further factorized in color space into a hard function, $H$, describing quanta of
high virtuality, and a soft function, $S$, accounting for soft gluon emission
into the central region, as follows \cite{BottsSt,KOS1,KOS2}:
\beqa
Q_c\frac{d\hat{\sigma}^{(\a)}}{dQ_c}\left(\hat{s},\hat{t},y_{JJ},\Delta y,\alpha_s(-\hat t)\right)
&=&
\nonumber \\
&& \hspace{-54mm} H_{IL}\left( \frac{\sqrt{-\hat{t}}}{\mu},
\sqrt{\hat{s}},\alpha_s(\mu^2) \right)  S_{LI} \left( \frac{Q_c}{\mu},y_{JJ},\Delta y\right). \nonumber \\
\label{factor}
\eeqa 
Here we can identify a hard scale, $\sqrt{-\hat{t}}$,  a soft scale, $Q_c$,
and a new factorization scale, $\mu$.
We sum over the indices $I$ and $L$, which label the possible color structures of
the hard interaction. 
For the scattering of valence quarks and antiquarks
these are just singlet and octet and, in a basis of $t$-channel projectors,
are given respectively by \cite{KOS2}
\beqa
c_1&=&\delta_{\e_A,\e_1}\delta_{\e_B,\e_2}\nonumber\\
c_2&=&-\frac{1}{2N_c}\delta_{\e_A,\e_1}\delta_{\e_B,\e_2}+\frac{1}{2}\delta_{\e_A,\e_B}\delta_{\e_1,\e_2}.
\label{eq:basqqbar}
\eeqa 

The soft matrix $S_{LI}$ in Eq.\ (\ref{factor}) coincides with the 
effective ``eikonal'' cross section, in which the
hard scattering is replaced by a product of recoilless Wilson lines \cite{BottsSt,KOS1,KOS2} 
in the directions
of the incoming partons and the outgoing jets.  
It starts to receive contributions at zeroth order in $\alpha_s$, where it is given by just a set of color traces,
\beq
S^{{(0)}}_{LI}=\left(\begin{array}{cc}
N^2_c & 0\\
0 & \frac{1}{4} \left( N^2_c-1 \right) 
\end{array} \right),
\label{softcol}
\eeq
with $N_c$ the number of colors.

$H_{IL}$, on the other hand, starts at the level of Born amplitudes. The contribution of $t$-channel gluon exchange,
which is pure color octet, is dominant, and we have $H^{(1)}_{IL}=\delta_{I2}\, \delta_{L2} \, \hat{\sigma}_t$,
where $\hat{\sigma}_t$ is the  $t$-channel 
partonic cross section, including 
the coupling $\alpha_s(-\hat{t})$. 

From the independence on $\mu$ of the left-hand side of Eq.\ (\ref{factor}), we immediately deduce
the evolution equation satisfied by the soft matrix $S_{LI}$,
\beqa 
\left(\mu\frac{\partial}{\partial\mu}+\beta(g)\frac{\partial}{{\partial}g}
\right)S_{LI}&=&
-(\Gamma_S^\dagger)_{LB}S_{BI} \nonumber \\
&& \hspace{2mm} -S_{LA}(\Gamma_S)_{AI}\, ,
\label{eq:resoft}
\eeqa
where $\Gamma_S(\alpha_s)$ is a soft anomalous dimension
matrix. The solution of this equation will enable us to resum all the leading logarithms of the soft scale $Q_c$.
In the singlet-octet basis of Eq.\ (\ref{eq:basqqbar}) the anomalous dimension matrix  is given by \cite{StOd1}
\beq
\Gamma_S\left(y_{JJ},\Delta y,\hat{\theta}\right)
=
\frac{\alpha_s}{4\pi} \left(
\begin{array}{cc}
\w+\g & -4\frac{C_F}{N_c}i\pi \\
-8i\pi & \w-\g
\end{array} \right),
\label{rapanodim}
\eeq
where the functions $\g$ and $\w$ are
\beqa
&&\g(\Delta y)=-2N_c \Delta y +2i\pi\frac{N_c^2-2}{N_c},
\label{defntg}
\\
&&\w(y_{JJ},
\Delta y,\hat{\theta}
)=\frac{N_c^2-1}{N_c} \nonumber \\
&& \times \left[ \ln \left( \frac{\cos(\hat{\theta})+
\tanh\left( \frac{\Delta y}{2}-y_{JJ}\right)}
{\cos(\hat{\theta})-\tanh\left( \frac{\Delta y}{2}-y_{JJ} \right)}\right) \right.
\nonumber \\
&&+ \left. \ln\left(\frac{\cos(\hat{\theta})-\tanh\left( -\frac{\Delta y}{2}-y_{JJ} 
\right)}
{\cos(\hat{\theta})
+\tanh\left( -\frac{\Delta y}{2}-y_{JJ} \right)}\right)\right] \nonumber \\ 
& & +\frac{2}{N_c}\Delta y-2i\pi\frac{N_c^2-2}{N_c}\, .
\label{defntw}
\eeqa

\section{Color Evolution}
\label{sec:facevoeiso}
It is natural to solve the evolution equation (\ref{eq:resoft}) in the basis of
the eigenvectors of $\Gamma_S$,
\beqa
e_1&=&\left(\begin{array}{c}
1\\
\frac{8\pi}{i}\left(\g-\frac{1}{\sqrt{N_c}} \y \right)^{-1} \\
\end{array} \right) \nonumber \\
e_2&=&\left(\begin{array}{c}
\frac{i}{8\pi}\left(\g+\frac{1}{\sqrt{N_c}} \y \right)\, , \\
1\\
\end{array} \right),
\label{eigenvectors}
\eeqa
where we define
\beq
\y(\Delta y)\equiv\sqrt{N_c \left[\g(\Delta y)\right]^2 -32C_F\pi^2}\, .
\label{ydef}
\eeq
The eigenvectors only depend on the geometry, through $\Delta y$.
In the limit of a wide central region $\Delta y \gg 1$ they
become pure color states, $e_1$ a singlet and $e_2$ an octet.
However, for a typical D0 geometry \cite{D0,D0fig}, $\Delta y=4$ at 
$\sqrt{S}=1800 \, {\rm GeV}$, we have in general color mixed states.
In the following we will refer to $e_1$ and $e_2$ as ``quasi-singlet'' and
``quasi-octet'' respectively, and we shall use Greek indices to identify the basis 
in which $\Gamma_S$ is diagonal \cite{KOS2}.

The eigenvalues of $\Gamma_S$ corresponding to the eigenvectors of Eq.\ (\ref{eigenvectors})
are given by  $\lambda_{\beta}=\alpha_s \hat \lambda_\beta+\cdots$, where we define
\beqa
\hat{\lambda}_1(y_{JJ},
\Delta y,\hat{\theta}
)&=&\frac{1}{2\pi}\left[\frac{1}{2}\, \w-
\frac{1}{2\sqrt{N_c}} \y  \right] 
\nonumber \\
\hat{\lambda}_2(y_{JJ},
\Delta y,\hat{\theta}
)&=&\frac{1}{2\pi}\left[\frac{1}{2}\w+
\frac{1}{2\sqrt{N_c}} \y \right] .
\label{eigenvalues}
\eeqa
The $\hat{ \lambda}_{\beta}$'s are in general complex. 
Over most of the $\hat{\theta}$, $y_{JJ}$ kinematical region
$\rm{Re} \hat{\lambda}_2 > \rm{Re} \hat{\lambda}_1$ as well as $|\hat{\lambda}_2| > |\hat{\lambda}_1|$.

The evolution equation (\ref{eq:resoft}) is easily solved when $\Gamma_S$ is diagonal,
to give the partonic cross section \cite{StOd1}
\beqa
\frac{d\hat{\sigma}^{(\a)}}{dQ_c \, d\cos \hat{\theta}}\left(\hat{s},\hat{t},y_{JJ},\Delta y,\alpha_s(-\hat t)\right)
&=& 
\nonumber \\
&& \hspace{-55mm} H^{(1)}_{\beta \gamma}\left( \Delta y,\sqrt{\hat{s}},\sqrt{-\hat{t}},
\alpha_s\left(-\hat{t}\right) \right) 
 S^{{(0)}}_{\gamma \beta} ( \Delta y ) \,  \nonumber \\
&& \hspace{-55mm} \times {E_{\gamma\beta} \over Q_c}\; 
\left[\ln\left({Q_c\over \Lambda}\right)\right]^{E_{\gamma\beta}-1}\;  
\left[ \ln \left( {\sqrt{-\hat{t}}\over \Lambda}\right)\right]^{-E_{\gamma\beta}}\, . \nonumber\\
\label{factor2}
\eeqa
Here $S^{{(0)}}_{\gamma \beta}$ and $H^{(1)}_{\beta \gamma}$ are obtained by transforming 
the corresponding quantities, defined in the basis (\ref{eq:basqqbar}), into the color eigenspace \cite{StOd1}.
The exponents  $E_{\gamma \beta}$ are given by
\beqa
&&E_{\gamma\beta}\left(y_{JJ},\hat{\theta},\Delta y \right)
=\frac{2\pi}{\beta_1}\, \left(\hat{ \lambda}^{*}_{\gamma} 
+\hat{ \lambda}_{\beta}  \right)\, ,
\label{expon}
\eeqa
where $\beta_1$ is the first coefficient in the expansion of the QCD
$\beta$-function, $\beta_1=\frac{11}{3}N_c-\frac{2}{3}n_f$.
The magnitude of the
quasi-singlet exponent $E_{11}$ is less than one for most kinematical
configurations, while the magnitude of the
quasi-octet exponent $E_{22}$ is always greater than one. As a consequence,
Eq.\ (\ref{expon}) shows that
the quasi-singlet component of the hard scattering is a decreasing function of $Q_c$,
dominant for $Q_c< 1 \, {\rm GeV}$ and formally divergent at $Q_c=\Lambda$,
whereas the quasi-octet component increases up to a maximum, until the inverse power behavior $1/Q_c$
takes over, causing a fast decrease.

\section{Numerical results }
\label{sec:results} 
\begin{figure}
\centerline{\epsffile{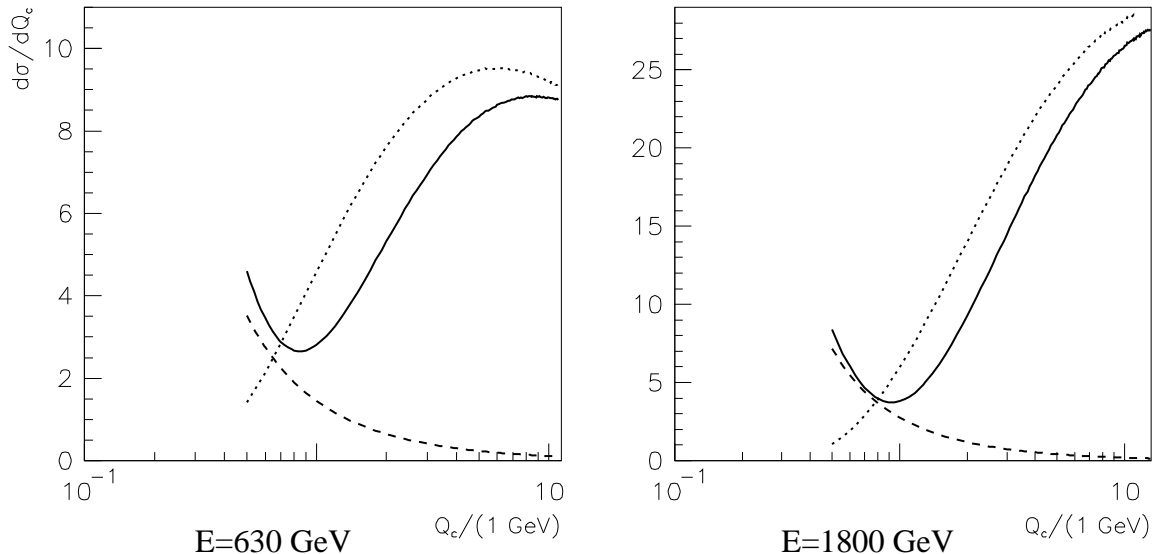}}
\caption[dum]{The cross section (solid line) and the contributions from
quasi-octet (dotted line) and quasi-singlet (dashed line), for
$\sqrt{S}=630$ GeV, $\Delta y=3.2$,
and  $\sqrt S =1800$ GeV, $\Delta y=4$, respectively.
Compare Fig.\ 1 of Ref.\ \cite{D0fig}.  Units are arbitrary.}
\label{fig4}
\end{figure}

By using in Eq.\ (\ref{crosssec}) the partonic cross section of Eq.\ (\ref{factor2})
and the set CTEQ4L of parton distribution functions \cite{cteq4}, and by performing the numerical integrations
with the routine VEGAS, we have obtained the results shown in Fig.\ \ref{fig4}.
These are to be compared with the experimental plot in Fig.\ 1 of Ref.\ \cite{D0fig},
showing the measured number of events as a function of
the number of
towers counted in the central region of the calorimeter, clearly related to our $Q_c$.
The shape of the experimental data is reproduced by our result. We emphasize that this result is perturbative.
For quantities independent of overall normalizations we have found similarity
with the data even at the quantitative level.
For example, the minimum-maximum
ratio of the cross section, is about $30 \%$ at $\sqrt{S}=630 \, {\rm GeV}$, 
and about $15 \%$ at $\sqrt{S}=1800 \, {\rm GeV}$, 
close to estimates we can make from the experimental data.
Also an  analog of the ``hard singlet fraction"
\cite{D0,CDF}, defined as the ratio of the area under the quasi-singlet curve
to the area under the overall curve,  is found to be about $5\%$ at 
$\sqrt{S}=630 \, {\rm GeV}$ and about $3 \%$ at $\sqrt{S}=1800\, {\rm GeV}$,
of the same order of magnitude, although somewhat 
higher, than the roughly $1\%$ found at the  Tevatron
using track or tower multiplicities.   
The precise origin of this discrepancy might be related to the non-perturbative part
of the survival probability \cite{GLM}, and remains to be explored.
The decrease of the singlet fraction as a function of the center of mass energy
is also in agreement with the experiment, although the trend we find is slower
than the measured one. 

\section{Summary}
We have computed by means of pQCD a dijet rapidity-gap cross section,
defined in terms of energy flow. We have shown that the numerical results
from this cross section qualitatively reproduce the behavior of the experimental data
of Ref.\ \cite{D0fig}.
Our analysis still needs to be refined by including the contributions
of gluons and sea-quarks \cite{StOd}.

\subsection*{Acknowledgments}
I would like to thank George Sterman, with whom the work I presented
was done, for his suggestions and constant encouragement.
I am also grateful to Jack Smith for his valuable advice 
in the implementation of the numerical simulation.  
This work was
supported in part by the National Science Foundation, grant PHY9722101.

\end{document}